\documentclass[a4paper]{iopart}

\usepackage{graphicx}
\usepackage{amssymb}
\usepackage{cite}

\usepackage{color}

\begin{document}

\title{Radiation from low-momentum zoom-whirl orbits}

\author{R. Gold and B. Br\"ugmann}

\address{Friedrich-Schiller University, 07743 Jena, Germany}

\ead{roman.gold@uni-jena.de, bernd.bruegmann@uni-jena.de}

\begin{abstract}
We study zoom-whirl behaviour of equal mass, non-spinning black hole
binaries in full general relativity. The magnitude of the linear
momentum of the initial data is fixed to that of a quasi-circular
orbit, and its direction is varied. We find a global maximum in
radiated energy for a configuration which completes roughly one
orbit. The radiated energy in this case exceeds the value of a
quasi-circular binary with the same momentum by $15\%$. The direction
parameter only requires minor tuning for the localization of the
maximum. There is non-trivial dependence of the energy radiated on
eccentricity (several local maxima and minima). Correlations with
orbital dynamics shortly before merger are discussed. While being
strongly gauge dependent, these findings are intuitive from a physical
point of view and support basic ideas about the efficiency of
gravitational radiation from a binary system.
\end{abstract}


\section{Introduction}

Compact binaries, especially black hole binaries, are expected to be
primary sources for gravitational wave astronomy. The orbit of a black
hole binary is often assumed to be a quasi-circular inspiral that arises
after circularization of the orbit due to the emission of gravitational
waves, since this is deemed the most likely astrophysical
scenario. However, as a matter of principle we should be prepared to
detect and recognize gravitational waves from all corners of the
parameter space of binaries.  So-called zoom-whirl orbits are of
particular interest because of their distinct gravitational wave
signature, and because in general eccentric orbits can be even
more efficient than the quasi-circular inspirals in converting
energy of a binary into gravitational waves. Zoom-whirl orbits of a
binary are orbits that consist of tight and fast revolutions (the
whirls) as well as a phase where the two objects move out to larger
distances and back in (the zooms).

The basic features of zoom-whirl orbits were first discussed in the
context of geodesics in a stationary black hole spacetime. This
describes the limit of a small test mass moving around a large central
black hole, see for
example~\cite{Cha83,martel-2004-69,levin_2009_79,grossman-2009-79,PhysRevD.50.3816,GlaKen02,hughes-2005-94}, some of which include radiation effects. 
The main question in full general relativity is how the classic,
well-known picture of zoom-whirl geodesics changes for binaries with
comparable masses in configurations where radiation damping becomes
significant. \textit{A priori} it is unclear how many whirls can be achieved by
fine-tuning the initial orbital parameters (arbitrarily many in the
case of geodesics) since the system is quickly losing energy and
angular momentum. If little tuning is involved, then zoom-whirl orbits
can be potential GW sources even for ground-based detectors, see for
example~\cite{Healy:2009zm,martel-1999-60,brown-2009,Wen:2002km}.

Recently, zoom-whirl orbits have become the subject of numerical
simulations in full general relativity.
In~\cite{PreKhu07}, Pretorius and Khurana presented the first
example of a whirl orbit of an equal mass black hole binary. 
In~\cite{sperhake-2008-78,hinder-2008-77,washik-2008-101}, several
examples for the transition from inspiral to plunge, radiated energy
and angular momentum have been studied.
In~\cite{Healy:2009zm}, up to three whirl phases have been found. 
The first study of binaries with non-vanishing spin can be found
in~\cite{healy-2009}, and the consequences for kicks have been studied
in~\cite{healy-2009-102}.
The focus in~\cite{Sperhake:2009jz,sperhake-2008-101,ShiOkaYam08} is on
high-energy collisions, where the kinetic energy becomes the dominant
contribution to the total mass of the system. 

In this paper we examine one particular type of zoom-whirl orbit for
equal masses and vanishing spin. The magnitude of the linear
momentum of the initial data is fixed to that of a quasi-circular
orbit at the given initial separation, and its direction is varied.
With our choice of initial data we probe a comparatively low-momentum
regime aiming at astrophysically reasonably realistic binaries.
Similar low-momentum cases have been studied in~\cite{Healy:2009zm}. 
The novel aspect of the present work is that we perform a detailed
investigation of how the energy radiated depends on the initial angle
and find a physical interpretation of this dependence.  Main result is
that there is not just one maximum in the radiated energy as a
function of the initial angle. Rather, several local maxima and minima
can be identified, which should be compared to the variations in the
mass and spin of the merger remnant noted in~\cite{healy-2009}. We
link these extrema to the motion of the punctures. Furthermore, for
this low-momentum case, the amount of fine-tuning required is small;
an angular resolution of about $1^\circ$ suffices.

The term `zoom-whirl' has obvious meaning when several or more whirl
orbits near an (unstable) quasi-circular orbit are found, but recent
literature colloquially refers to orbits with one or two whirls also
as `zoom-whirl' orbits. The physics behind the whirls is precession,
and any eccentric orbit has some amount of precession. This begs the
question at what point large precession becomes a whirl. We refer to
an orbit as a zoom-whirl orbit if at least one whirl orbit is
completed before the next zoom, i.e.\ if from one apoastron of a zoom
to the next an additional angle of $2\pi$ or more is traversed.
However, in order to distinguish large precession on the order of
$\pi$ that occurs for certain black hole mergers (as discussed in this
paper) from perturbative precession effects on the order of
arcseconds, we refer to precession effects on the order of $\pi$ also
as partial or fractional whirls.

\section{Numerical Computations}

We perform a parameter study of equal mass, nonspinning black hole
binaries by varying the direction of the initial momentum (see below).
We use the BAM-code~\cite{BruTicJan03,BruGonHan06} to solve the full
Einstein equations in vacuum. The problem is cast as a
Cauchy-initial-value problem by foliating the four dimensional
spacetime into families of three-dimensional hypersurfaces labeled by
a coordinate time $t$. Initial data for the initial hypersurface are
obtained by the puncture method using a pseudo-spectral
code~\cite{BraBru97,AnsBruTic04} assuming the spacetime being
conformally flat. The computational method uses adaptive mesh
refinement (typically $7-10$ levels) together with Berger-Oliger
timestepping. We use the $\chi$-variant of the
moving-puncture~\cite{CamLouMar05,BakCenCho05} version of the
Baumgarte-Shapiro-Shibata-Nakamura (BSSN)
formulation~\cite{ShiNak95,BauSha98}. The runs are performed with
sixth-order finite differencing in space~\cite{HusGonHan07} and a
fourth-order Runge-Kutta method. For the purpose of wave extraction
we use an implementation of the Newman-Penrose formalism which is also
fourth order accurate. We can demonstrate consistent, overall fourth-order
convergence in the 22-mode of $r\Psi_4$ and the radiated energy
$E_{rad}$ with typical relative errors of $1-2\%$ for $E_{rad}$ and
slightly less than $1\%$ for the 22-mode. The error due to a finite
extraction radius was estimated by extracting the waves at different
radii $r=60M,80M,100M$ (where $M$ is the total puncture mass) and
computing the deviations from the expected $1/r$ fall-off. This error
turned out to be roughly of the same order as the discretization error
mentioned above.

Two punctures of mass $m$ are placed on the $x$-axis at $x_1=-10M$ and
$x_2=+10M$, respectively, where $M=2m$, so that the initial coordinate
separation is $D=20M$. The magnitude $P$ of the linear momentum
$\vec{P}$ of each puncture is set equal to the value for quasi-circular
orbits, $P=P_{qc}=0.061747M$, using the methods of~\cite{WalBruMue09}.
We systematically vary the direction of $\vec{P}$, which implies
corresponding changes in the initial angular momentum and also the
eccentricity of the orbit. This is parameterized by a shooting angle
$\Theta$ defined as the angle spanned by the initial linear momentum
vector $\vec{P}$ and the $x$-axis.  Alternatively, we could e.g. vary
the size of the initial momentum, which we leave for a future
investigation. Varying the values of $\Theta$ from $\Theta=30^{\circ}$
to $\Theta=53^{\circ}$ we compute $E_{rad}$ and investigate how the
energy might be related to the trajectories of the punctures. This study
involves 24 binary black hole evolutions, plus convergence runs.

\section{Results}

\begin{figure}[p]
  \centerline{\resizebox{10cm}{!}{
\includegraphics{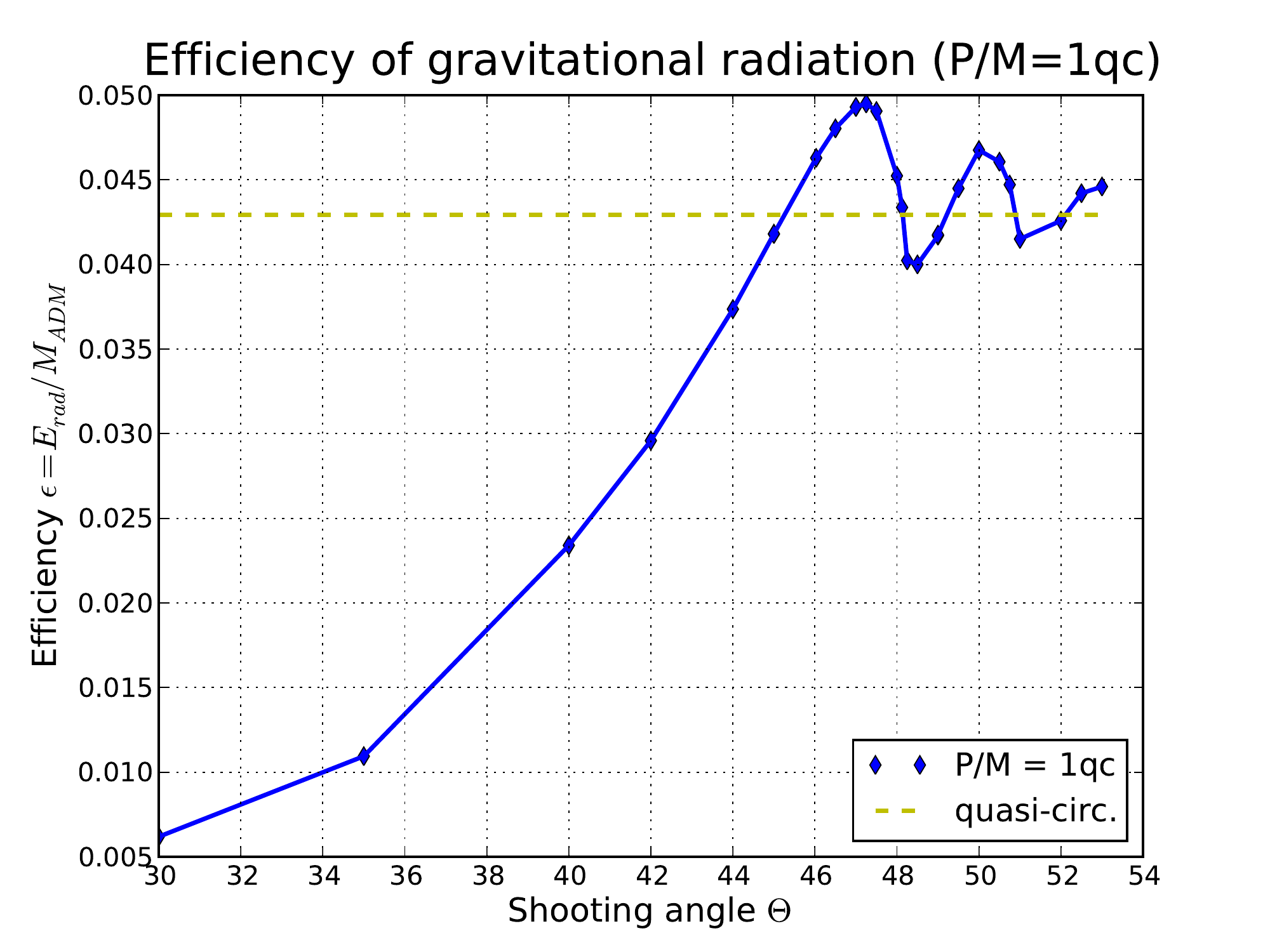}}}
\caption{
Efficiency of gravitational radiation as a function of the shooting
angle $\Theta$. For the $y$-axis we normalize $E_{rad}$ by the
ADM-mass of the initial time slice. The maximum at
$\Theta\approx47^{\circ}$ corresponds to an initial orbital angular
momentum $L=D P \sin(\Theta)=0.88M^2$.
}
\label{fig:Erad}
\end{figure}

\begin{figure}
\centerline{\resizebox{10cm}{!}{
\includegraphics{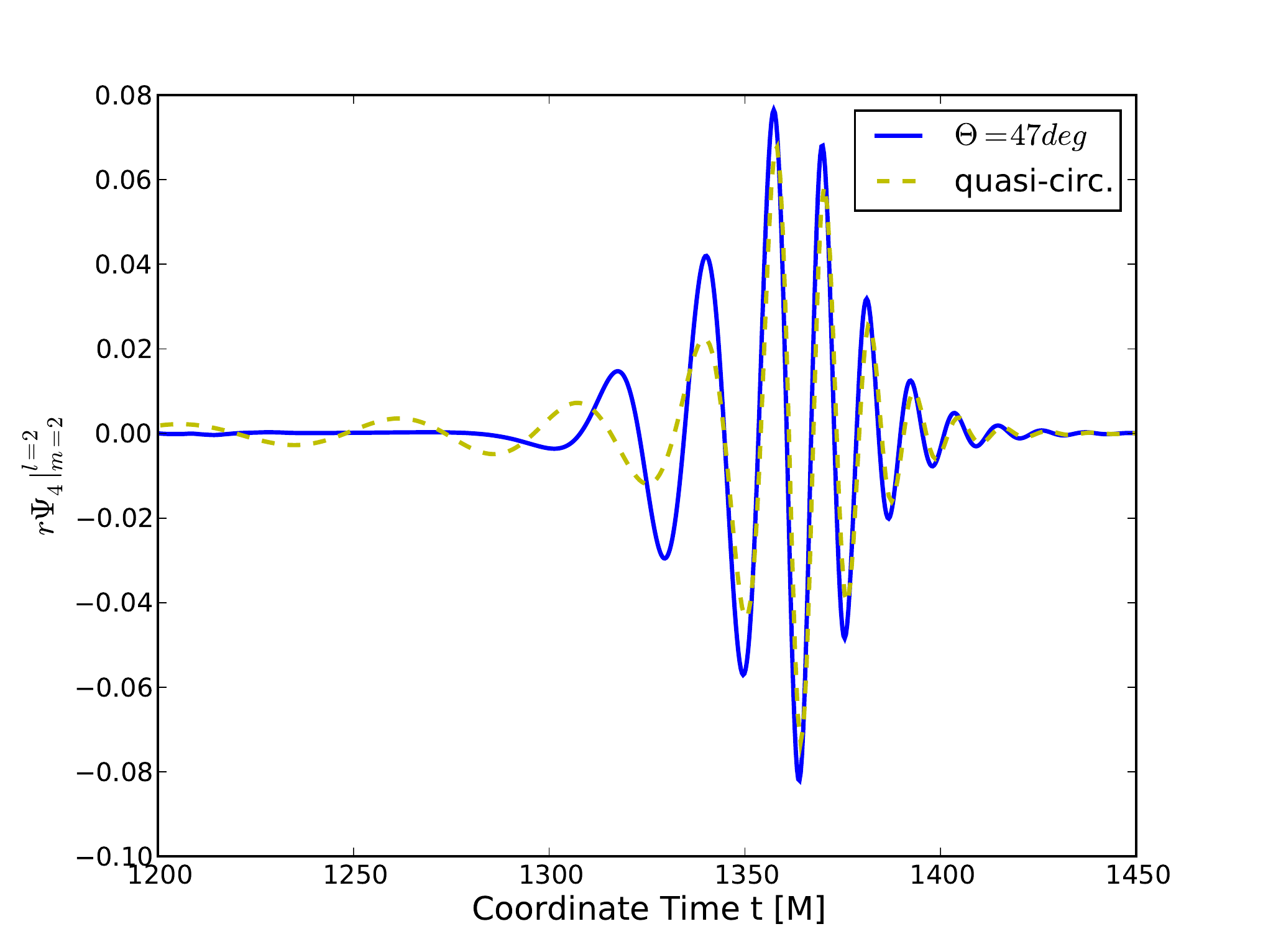}}}
\caption{
Comparison of 22-modes of the most efficient case ($\Theta=47^{\circ}$) and
the quasi-circular ($\Theta=90^{\circ}$) case. The late-time, ring-down
waveforms are quite similar. Differences are visible before the merger
where the amplitude for the $\Theta=47^{\circ}$ case is up to twice as
large as for the quasi-circular binary, and in the inspiral part where
the quasi-circular orbit results in more radiation.  
}
\label{fig:Waveforms}
\end{figure}

\begin{figure}[p]
  \centerline{\resizebox{10cm}{!}{
\includegraphics{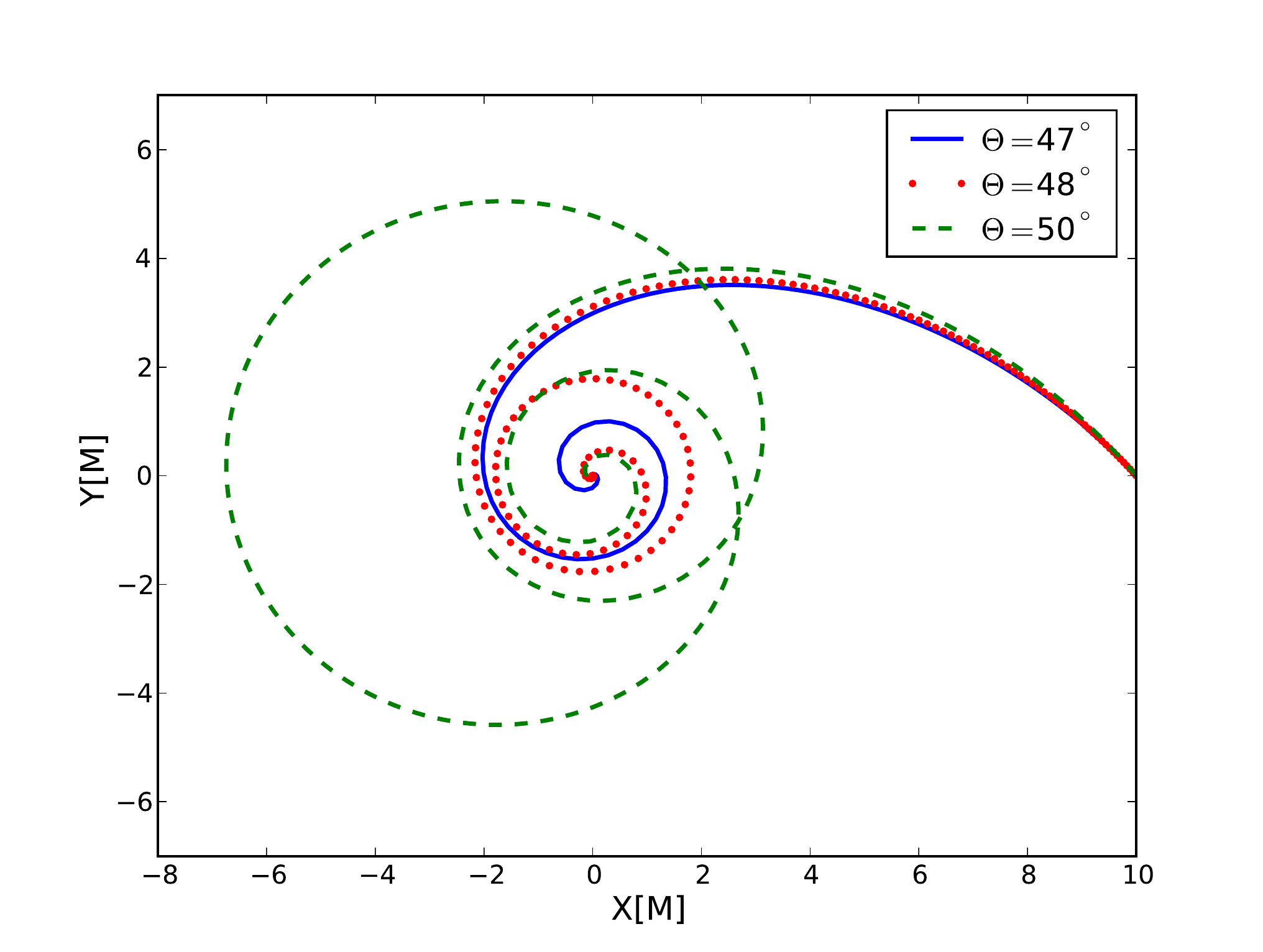}}}
\caption{ Trajectories of three evolutions corresponding to one
  minimum and two maxima in $E_{rad}$ (compare $\Theta$ values in the
  legend with $x$-axis in figure~\ref{fig:Erad}). The solid line
  represents an inspiral without noticeable whirl or zoom, the dotted
  line shows a complete whirl before the plunge, and the dashed line
  starts with a fractional whirl (less than a full whirl orbit but
  strong precession), followed by a zoom phase and the plunge.  Only
  one puncture trajectory for each evolution is shown for clearity.  }
\label{fig:Orbits}
\end{figure}

We analyze the dependence of the radiated energy $E_{rad}$ on the
shooting angle $\Theta$ and observe new features for eccentric,
equal-mass and non-spinning binary black holes including zoom-whirl
behaviour. The results on radiated energy are summarized in
figure~\ref{fig:Erad}. Figure~\ref{fig:Waveforms} compares two of the
waveforms, and the full orbital trajectories of three representative
cases are shown in figure~\ref{fig:Orbits}.

First, we note that for the size of the linear momentum considered here
eccentric black hole binaries can be significantly ($15\%$), but not
drastically more efficient than quasi-circular binaries in converting
initial energy into outgoing gravitational radiation. This can also be
seen from figure~\ref{fig:Waveforms}, where we plot the $l=2,m=2$ mode of
$r\mathcal{R}e(\Psi_4)$ for the $\Theta=47^{\circ}$ run against a
quasi-circular waveform. Especially at the onset of merger the amplitude
of the eccentric waveform is significantly larger than in the
quasi-circular case. This will be further explored in an upcoming work
in the context of larger initial momenta. Second, for $\Theta$ in the
range of $45^{\circ}-53^{\circ}$, $E_{rad}$ fluctuates between
$0.04M_{ADM}$ and $0.05M_{ADM}$.  The oscillations are much larger than
the estimated relative numerical error of $1-2$\%, and are therefore
considered a feature of the physics rather than a numerical artefact.
Analogous variations have been observed in the mass of
the merger remnant as well as in the final spin in~\cite{healy-2009}. 
However, these oscillatory features have not been observed by
\cite{Sperhake:2009jz,hinder-2008-77}, while the behaviour for low 
$\Theta$ looks very similar to \cite{sperhake-2008-78}. In the
following we will explore the origin of the above findings in more
detail.

\begin{figure}[p]
  \centerline{\resizebox{10cm}{!}{
\includegraphics{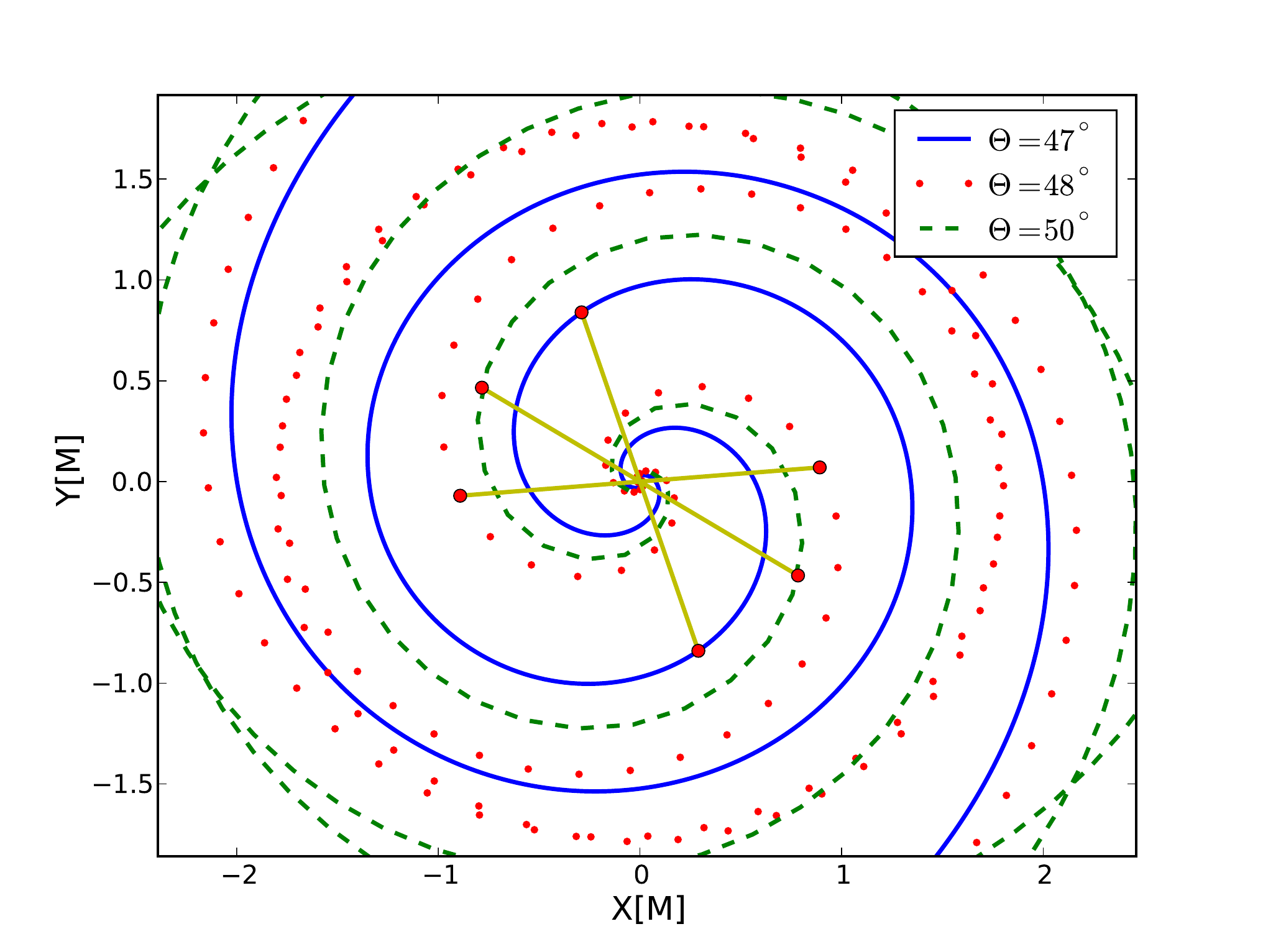}}}
\caption{ Zoom into the inner part of three orbital evolutions
  corresponding to one minimum and two maxima in $E_{rad}$ (compare
  $\Theta$ values in the legend with $x$-axis in
  figure~\ref{fig:Erad}). The markers denote the onset of merger. A
  maximum in $E_{rad}$ occurs when the coordinate separation vector
  $\vec{S}$ (straight line) is closest to being orthogonal to the
  tangent vector of the puncture tracks at the time of merger.  }
\label{fig:TangentialMergers}
\end{figure}

\begin{figure}[p]
\centerline{\resizebox{10cm}{!}{
\includegraphics{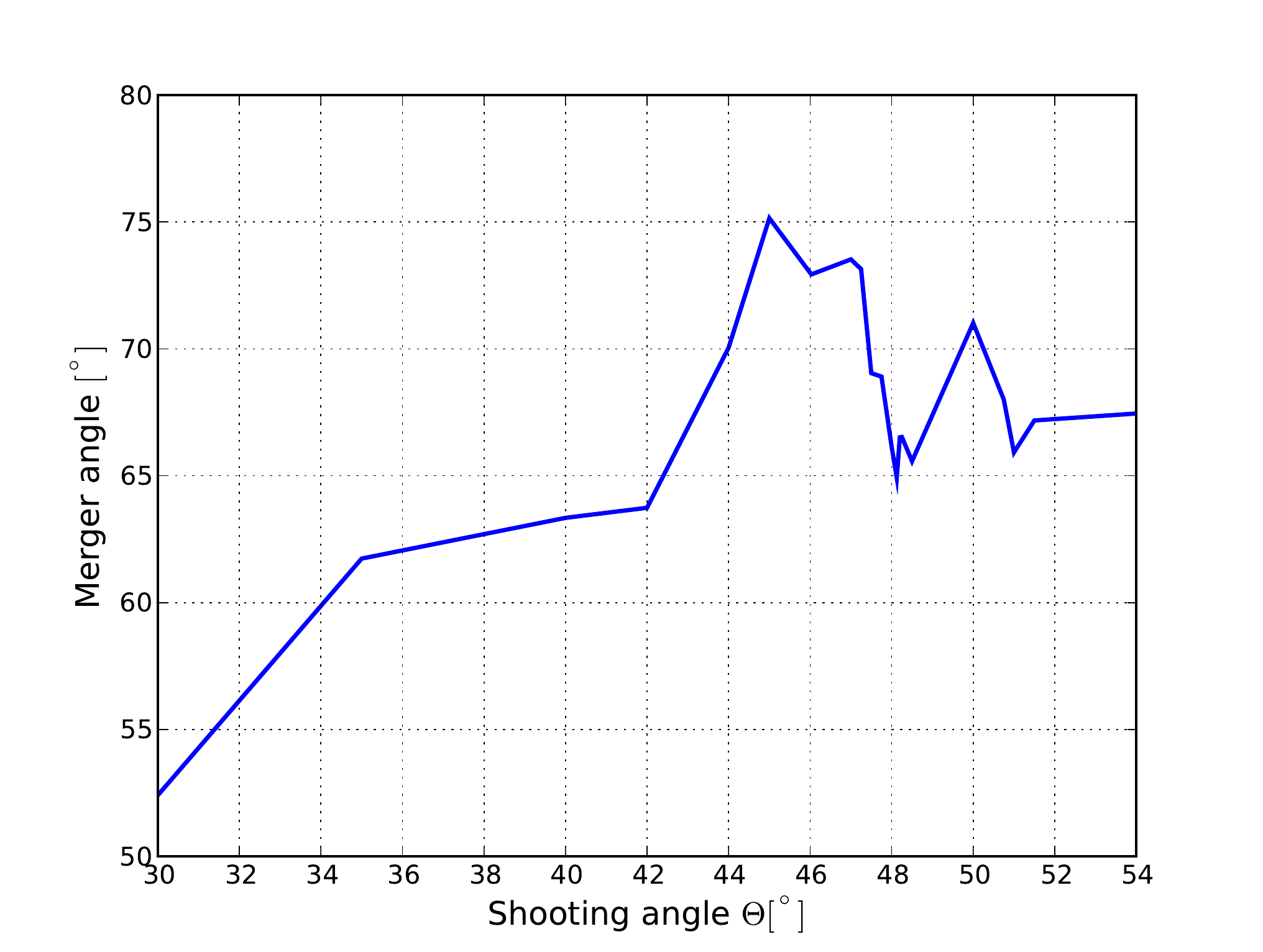}}}
\caption{The angle between the tangent vector of the puncture orbit
  and the separation vector $D$ as a function of the initial shooting
  angle.  This plot looks similar in its general shape to the plot for
  the radiated energy in figure~\ref{fig:Erad}. The locations of the
  maxima and minima basically coincide.  }
\label{fig:MergerAngle}
\end{figure}

We find strong evidence that the local minima and maxima are linked to
different types of dynamics near the merger. In order to make a more
precise statement we define the coordinate separation vector $\vec{S}$
pointing from one puncture to the other with length
$|\vec{S}(t)|=D(t)$ for all $t$, and we denote the time when a common
horizon is formed by $t_{merger}$. We depict the following observation
in figure~\ref{fig:TangentialMergers}. Whenever the tangent vector of
the puncture orbit is closest to being orthogonal to
$\vec{S}(t=t_{merger})$ then the corresponding orbit turns out to
maximize the radiated energy compared to other evolutions in the
neighbourhood in parameter space. In contrast, those orbits whose
tangent vectors point furthest inwards at the onset of merger (e.g.\
$\Theta=48^{\circ}$, dotted line in figure~\ref{fig:TangentialMergers})
minimize the radiated energy. Figure~\ref{fig:MergerAngle} shows the
angle at merger time for a number of different runs. The locations of
the maxima and minima in figure~\ref{fig:MergerAngle} mimic the
dependence of the energy on the shooting angle shown in
figure~\ref{fig:Erad}.

A physical interpretation of this effect is intrinsically complicated
due to the high degree of gauge dependence in this highly nonlinear
region in spacetime. Nevertheless, the above-mentioned effect might
originate from the fact that initial eccentricities within a certain
range can maximize or minimize the time the binary spends in the
radiatively most efficient configuration, namely small separations and
high velocities. It is rather intuitive that orbits with radial velocity
$v_r \approx 0$ shortly before merger (e.g.\ $\Theta=47^{\circ}$, solid
line in figure~\ref{fig:TangentialMergers}) spend comparatively more time
in a radiatively-efficient zone than in the case where $v_r$ is
larger. A large $v_r$ leads to a rapid rush through this crucial zone
leaving the binary less time to radiate.

\begin{figure}[p]
\centerline{\resizebox{10cm}{!}{
\includegraphics{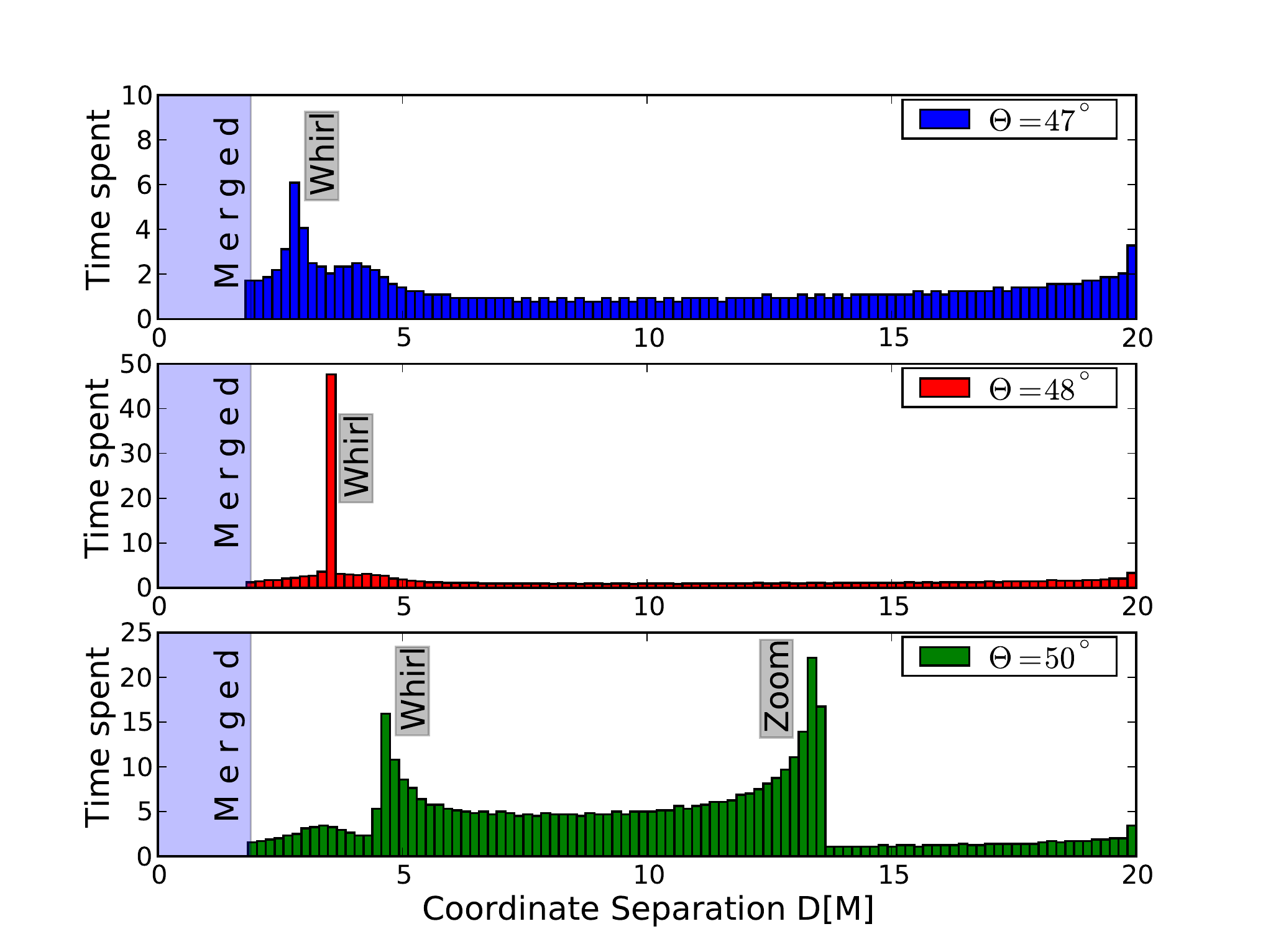}}}
\caption{
Another comparison regarding efficiency using a histogram of $D$. Shown
is how much time the binary has spent at $D\pm0.2$. The histograms
include data up until the merger time $t_{merger}$ which read $137M$,
$184M$ and $432M$, upper to lower, respectively. Less efficient binaries
(middle panel) spent more time at a larger separation whereas efficient
ones (upper panel) at lower separations. In the lower panel one can see
that both whirl and zooms cause peaks in this diagram.}
\label{fig:Hist}
\end{figure}

This interpretation is further supported by the data shown in
figure~\ref{fig:Hist}. In this plot we compute a histogram of $D=D(t)$,
i.e.\ we count the number of time steps during which the coordinate
separation of the binary lies within each bin of width $\Delta D$, thereby 
visualizing how much time the binary spends at a given 
separation $D$.
First, we observe that we can define the radius for
innermost unstable circular orbits rather well for our gauge-choice,
because in this plot they show up as a sharp and strong peak. Second,
the comparison between the upper and lower panel confirm our
interpretation, that more efficient runs spend more time in a
radiatively-efficient zone than less efficient binaries. In this plot
the emission zone corresponds to a band bounded on the left by
$D_{Merger}$, which is defined as the coordinate separation at which a
common horizon forms. We find $D_{Merger}=(1.8\pm0.1)M$.
We want to check to what degree the above hypothesis
withstands the issue of gauge dependence in an upcoming work. The idea
is to further investigate the momentum dependence (which the gauge is
sensitive to) and work out other diagnostics in order to clarify
whether or not such arguments carry over to more general
scenarios. 

Concerning the phenomenology of the orbital dynamics and the radiated
energy, we should emphasize that the maximum in radiated energy
corresponds to an evolution which only completes about $1.3$ orbits
before merger, i.e.\ the orbit is highly eccentric. (A
3PN~\cite{memmesheimer-2005-71} estimate for $\Theta=48^{\circ}$ gives
$e\approx 0.8$.)
For $48^\circ\leq\Theta\leq48.5^{\circ}$, i.e.\ near one of the minima
and not at one of the maxima, the orbits exhibit a `whirl' phase
slightly before merger.
This whirl phase does not seem to be the most relevant moment for wave
emission since we find that this almost circular motion happens at a
much larger radius (which depends inversely on the eccentricity) than
the coordinate separation at merger (see figure~\ref{fig:Hist}).
This argument is supported by the observation that the orbits which
manage to whirl, but do not exhibit a zoom {\em minimize} $E_{rad}$. For
$\Theta=48.5^{\circ}$ the whirl phase is followed by a very small
`zoom'. Above that value for $\Theta$ the whirl phase
becomes shorter and the zooms larger (note the prominent zoom of the
$\Theta=50^{\circ}$ run (dashed line) in figure~\ref{fig:Orbits}), and for
$\Theta>52^{\circ}$ there are multiple subsequent close encounters (or
fractional whirls) before the final merger.

In the strong field regime, the precise definition of eccentricity and
precession is problematic. Furthermore, the precession of the orbit is
no longer proportional to the eccentricity. 
For a large range of shooting angles (eccentricities) the precession
becomes a significant fraction of $2\pi$ and for a small range it even
exceeds $2\pi$. This is one of the main results of our (and other)
investigations. Prior to numerical simulations it was not clear
whether zoom-whirl behaviour exists at all in non-spinning, equal mass
binaries.
Indeed, as we demonstrate, the range where full whirls occur is very
limited, and the more typical phenomenology is rather governed by a
strong form of precession. The transition from close encounters to
zoom-whirls is completely smooth, and we especially want to highlight
that there is no significant fine-tuning involved beyond about
$1^\circ$ in the shooting angle.

The expectation that a high degree of fine-tuning may be required
originates with the search for geodesics that complete a large number
of whirls. These are typically found near an unstable circular orbit
(inside the radius of the innermost stable circular orbit), which
suggests that due to the instability a high degree of fine-tuning is
required. Also, the more time is spent in the whirl, the stronger the
dependence on the initial parameters.
In our case, only about one full whirl is found, so the sensitivity on
the initial angle should be modest.  Furthermore, although the whirl
occurs near an unstable orbit with a diameter of $3.5M$
(figure~\ref{fig:Orbits}), the degree of instability seems to be
sufficiently mild to require only a small amount of fine-tuning.

These findings are very different from the high energy
case~\cite{PreKhu07,Sperhake:2009jz}, where furthermore evidence for
critical scaling around the so-called threshold of immediate merger
has been found. In this regime highly fine-tuned initial conditions
are necessary. We plan to investigate larger momenta in future work.

\ack

The authors wish to thank 
M. Campanelli, L. Gergely, J. Grigsby, D. Hilditch, J. Levin,
D. M\"uller, G. Sch\"afer, D. Shoemaker and U. Sperhake
for discussions.
This work was supported in part by DFG SFB/Transregio~7
and DFG Research Training Group 1523. The computations were carried out
on the HLRB2 system at the LRZ in Garching.


\section*{References}
\bibliographystyle{unsrt}
\bibliography{refs,refsextra}


\end{document}